\documentclass{sig-alternate-2013}
\newfont{\mycrnotice}{ptmr8t at 7pt}
\newfont{\myconfname}{ptmri8t at 7pt}
\let\crnotice\mycrnotice%
\let\confname\myconfname
\usepackage{times}
\usepackage{helvet}
\usepackage{courier}
\usepackage{graphicx}
\usepackage{url}
\usepackage{algorithm}
\usepackage{algpseudocode}
\usepackage{amsmath}
\usepackage{amssymb}
\usepackage{amsfonts}
\usepackage{bbm}
\usepackage{booktabs,multirow}
\usepackage{subfig}
\usepackage{flushend}
\usepackage{multirow}
\usepackage{float}
\usepackage{color}
\usepackage[shortlabels]{enumitem}

\def\pprw{8.5in}
\def\pprh{11in}

\permission{Permission to make digital or hard copies of all or part of this work for personal or classroom use is granted without fee provided that copies are not made or distributed for profit or commercial advantage and that copies bear this notice and the full citation on the first page. Copyrights for components of this work owned by others than ACM must be honored. Abstracting with credit is permitted. To copy otherwise, or republish, to post on servers or to redistribute to lists, requires prior specific permission and/or a fee. Request permissions from permissions@acm.org.}
\conferenceinfo{RecSysChallenge'14,}{October 10, 2014, Foster City, CA, USA.}
\copyrightetc{Copyright 2014 ACM \the\acmcopyr}
\crdata{Copyright is held by the owner/author(s).\\Publication rights licensed to ACM.
ACM 978-1-4503-3188-3/14/10...\$15.00
http://dx.doi.org/10.1145/2668067.2668072}

\begin{document}

\title{Predicting User Engagement in Twitter\\ with Collaborative Ranking}

\numberofauthors{1} 
%
\author{
%
%
\alignauthor Ernesto Diaz-Aviles, Hoang Thanh Lam, Fabio Pinelli, Stefano Braghin,\\ Yiannis Gkoufas, Michele Berlingerio, and Francesco Calabrese\\
\affaddr{$ $} \\
\affaddr{IBM Research -- Ireland}\\
\affaddr{$ $} \\
\email{\{\mbox{e.diaz-aviles}, t.l.hoang, fabiopin, stefanob, yiannisg, mberling, fcalabre\}@ie.ibm.com}
%
%
%
%
%
%
%
%
%
%
}
\maketitle
\begin{abstract}
Collaborative Filtering (CF) is a core component of popular web-based services such as Amazon, YouTube, Netflix, and Twitter. Most applications use CF to recommend a small set of items to the user. For instance, YouTube presents to a user a list of top-n videos she would likely watch next based on her rating and viewing history. Current methods of CF evaluation have been focused on assessing the quality of a predicted rating or the ranking performance for top-n recommended items. However, restricting the recommender system evaluation to these two aspects is rather limiting and neglects other dimensions that could better characterize a well-perceived recommendation. In this paper, instead of optimizing rating or top-n recommendation, we focus on the task of predicting which items generate the highest user engagement. In particular, we use Twitter as our testbed and cast the problem as a Collaborative Ranking task where the rich features extracted from the metadata of the tweets help to complement the transaction information limited to user ids, item ids, ratings and timestamps. We learn a scoring function that directly optimizes the user engagement in terms of nDCG@10 on the predicted ranking. Experiments conducted on an extended version of the MovieTweetings dataset, released as part of the RecSys Challenge 2014, show the effectiveness of our approach.
\end{abstract}


\section{Introduction}
\label{sec:introduction}
Collaborative Filtering (CF) is an important method that helps users to deal with information overload by making personalized predictions of items of interest based on the preference information of many users. CF is also important to web-based services such as Netflix, YouTube, Amazon, and Twitter, which are able to improve customer satisfaction and increase revenue based on personalized recommendations.

Two fundamental tasks of CF are rating prediction and top-n recommendation. In particular, the latter can be cast as a ranking problem whose goal is to define an ordering among items (ie.g., Web pages, documents, news articles, Web sites, CDs, books, or movies) placing
relevant ones in higher positions of the retrieved list. Despite the advances in recent years of supervised learning-to-rank 
methods~\cite{2011_letor}, with very few exceptions (e.g.,~\cite{Diaz-Aviles:2012:SRR:2365952.2366001,Balakrishnan:2012:CR:2124295.2124314}) their application to CF has gained relatively little attention. A potential reason is that learning-to-rank approaches usually rely upon pre-engineered features that characterize the items to be ranked. Such a set of features can be difficult to compute or maintain, and kNN methods~\cite{Sarwar:2001:ICF:371920.372071} or matrix factorization methods, both for ranking (e.g.,~\cite{Diaz-Aviles:2012:RTR:2365952.2365968,Rendle:2009:BBP:1795114.1795167})  and regression (e.g.,~\cite{2006_funk,Koren:2008:FMN:1401890.1401944}), have been the preferred approaches for CF when only transaction information in the form of user-item pairs is available.

The core contribution of this work is a collaborative ranking approach for user engagement prediction in Twitter. We make use of the rich Twitter metadata to extract features that characterize user-item-tweet interactions. Such features help us to create a training dataset amenable to a learning-to-rank task, which allows us to leverage one of the many methods capable to directly optimize Information Retrieval (IR) metrics such as nDCG@10.

\subsection*{Collaborative Ranking Framework}
We cast the problem of user engagement prediction in Twitter as a Learning to Rank task~\cite{2011_letor}. In learning (training), a collection of users and their corresponding tweets are given. Furthermore, the \emph{engagement} (i.e., relevance judgments or labels) of the tweet with respect to the user are also provided. The engagement for a user-item-tweet triple, $(u,i,d)$, is defined as the expected user interaction, which is expressed in terms of sum of retweets and favorite counts:
$$
\text{engagement}(u,i,d) = \text{retweets}(u,i,d) + \text{favorites}(u,i,d) \; .
$$

Based on the value of engagement is possible to represent ranks of tweets per user. The objective of learning is
to construct a ranking model, e.g., a ranking function, that achieves the best result on test data in the sense of optimization of a performance measure, e.g., nDCG@10~\cite{2011_ricBaeza}.

In the recommendation phase (test), given a unseen user, the learned ranking function is
applied, returning a ranked  list of tweets in descending order of their
relevance scores. Suppose that $U=\{u_1,\cdots,u_{|U|}\}$ is the set of users. Users express their preferences over items using tweets, let $I$ be the set of such items, e.g., the user can comment about a movie or video or give an explicit rating in a tweet, in this case the movie item $i$ belong to set $I$. Furthermore, let $D=\{d_{1},\cdots,d_{|D|}\}$ be the set of tweets, then the training set is created as a set of user-item-tweet triples, $(u,i,d) \in U \times I \times D $, upon
which a relevance judgment (e.g., a label) indicating the user engagement for $u$ over item $i$ expressed in tweet $d$ is associated to the triple. 

Suppose that $Y=\{y_{1},\cdots,y_{|Y|}\}$ is the set of labels and $y_{uid} \in Y$ denotes the
label of user-item-document triple. A feature vector $\phi(u,i,d)$ is
created from each triple $(u,i,d)$. The training set is denoted by $S=\phi(u,i,d)_k,
y_{{uid}_k}\}_{k=1}^{m}$, where $m$ is the number of training examples, i.e., the number of observed triples. The ranking model is a real valued function of features:
\begin{equation}
   f(u,i,d, \Theta) = h(\phi(u,i,d); \theta) + b_{uid}
 \label{eq:rankingFunction}
\end{equation}
where $\Theta = \{\theta, b_{uid}\}$ is the set of free parameters to be learned. Here $h(\phi(u,i,d); \theta)$ is a scoring function on the user-item-tweet feature vectors $\phi(u,i,d)$, and $b_{uid}$ is a bias term. 

In ranking for user $u$, the model associates a personalized score to each of the tweets $d$ as their degree of relevance with respect to the user $u$ using $f(u,i,d, \Theta)$, and sorts the tweets based on their scores, e.g., in decreasing order. 

\section{Our Approach}
\label{sec:methods}
The main idea behind our approach is to transform the user engagement prediction problem into a collaborative ranking one and
then learn the ranking function using a learning-to-rank method~\cite{2011_letor}. The collaborative ranking task can be
placed into the learning-to-rank framework by noting that the users correspond to queries and tweets to documents. For each user the observed engagement indicates the relevance of the corresponding tweets with respect to that user and can be used to train the ranking function. The basic steps of our approach are summarized in Procedure~\ref{alg:cr_for_user_engagement}.

\floatname{algorithm}{Procedure}
\begin{algorithm}[!h]
\caption{Collaborative Ranking for User Engagement}
\renewcommand{\algorithmicrequire}{\textbf{Input:}}
\renewcommand{\algorithmicensure}{\textbf{Output:}}
\begin{algorithmic}[1]
\Require Training set $S=\{\phi(u,i,d)_k, y_{{uid}_k}\}_{k=1}^{m}$
\Ensure Ranking function $f(u,i,d, \Theta)$
\State Scale and normalize feature vectors $\phi(u,i,d)$
\State Remove user outliers
\State Learn a ranking function 
\Statex $\;\;\;\;\;$ $f(u,d,i, \Theta) = h(\phi(u,i,d); \theta) + b_{uid}$ 
\Statex by optimizing nDCG@10 directly
\State\Return $f(u,i,d, \Theta)$
\end{algorithmic}
\label{alg:cr_for_user_engagement}
\end{algorithm}

In the rest of the section, we present our feature extraction approach followed by the details on how the ranking function is learned.

\subsection*{Feature Extraction}
We propose to use the user-item-tweet interactions and the corresponding metadata of the tweets as main source to extract the feature vectors. The dataset used in this work corresponds to the \emph{MovieTweetings Extended}, released as part of the RecSys Challenge~\cite{2014_recsys_challenge_1, 2014_recsys_challenge_2}, 2014. The dataset is an extended version of the MovieTweetings dataset~\cite{Dooms13crowdrec}. The data includes the ratings extracted from users of the Internet Movie Database (IMDb) iOS app\footnote{\textbf{IMDb iOS app: }\url{http://www.imdb.com/apps/ios/}} that rate movies and share the rating on Twitter. MovieTweetings dataset is built by extracting information of the tweets on a daily basis by using Twitter API~\cite{Dooms13crowdrec}. The original MovieTweetings contains only rating information, i.e., (user, item, rating, timestamps) tuples. The extended version released for the challenge also includes all metadata of tweets as provided by Twitter API.\footnote{\textbf{Twitter API: }\url{https://dev.twitter.com/}}

We extract the following features $F$ to form the $\phi(u,i,d)$ feature vectors for each user-item-tweet triple.

\begin{enumerate}[F\arabic*:]
\item User \emph{rating}, $F1 = r_{uid}$, given to the movie item $i$ as expressed in tweet $d$.
\item Deviation of user rating with respect to the median of previous user ratings, i.e., $F2 = r_{uid} - \tilde{r}_{u}$
\item Average user engagement from her history, i.e., $$F3 = \overline{engagement}(u)^{0.5}$$
\item Boolean indicator that takes the value of 1 if the average user engagement is greater than 0, and 0 otherwise, i.e., $$F4=\mathbbm{1}[\overline{engagement}(u) > 0]\;,$$ where $\mathbbm{1}[a]$ is one if $a$ is true.
\item Average rating per user, $F5 = \bar{r_u}$
\item Ratio of number of user friends to the number of her followers, i.e., $$F6 = \left(\frac{\#friends(u)}{\#followers(u)}\right)^{0.5}$$
\item User tweet count: $$F7=\#tweets(u)^{0.5}$$
\item Average user engagement for a given movie item $i$ , i.e., $$F8=\overline{engagement}(i)^{0.5}$$
\item Boolean indicator that takes the value of 1 if the average user engagement for item $i$ is greater than 0, and 0 otherwise, i.e., $$F9=\mathbbm{1}[\overline{engagement}(i) > 0]$$
\item Average rating per item, $F10 = \bar{r_i}$
\item Average ratio of number of user friends to the number of her followers aggregated over the movie item $i$:
$$
F11 = \left( \frac{1}{K} \sum_{u \in U_i }^K \frac{\#friends(u)}{\#followers(u)} \right)^{0.5}
$$
where $U_i := \{u \in U | (u,j,d) \in S \land j=i \}$ and $K=|U_i|$ .
\item  Average of user tweet counts aggregated per item $i$ :
$$
F12 = \left( \frac{1}{K} \sum_{u \in U_i }^K \#tweets(u) \right)^{0.5}
$$
\item User mentions: boolean indicator that takes the value of 1 if the tweet contains any mention, i.e., any Twitter update that contains ``@username´´ anywhere in the body of the Tweet, and 0 otherwise, i.e., $$F13=\mathbbm{1}[has\_mention(d)]\;$$
\item Tweet is a retweet. From the tweet metadata field \break\texttt{retweeted\_status} we can infer if the current tweet $d$ is actually a retweet and use an boolean indicator to capture this information:  $$F14 = \mathbbm{1}[is\_retweet(d)]$$
\item The same field \texttt{retweeted\_status} for $d$ also includes the tweet id ($tweet\_id(d_o)$) of the original tweet, if such tweet $d_o$ is present in the dataset we know that it received a non-zero engagement, $F15$ represents this additional information for $d_o$: $$F15_{\phi(u,i,d_o)} = \mathbbm{1}[is\_retweet\_of(d, d_o)] \quad ,$$ where $is\_retweet\_of(d, d_o)$ is true if $d$ is $d_o$'s retweet.
\item The frequency of observed engagement (i.e., retweet count) extracted per item from the \texttt{retweeted\_status} field: $$F16 = \sum_{d \in is\_retweet(D)} \mathbbm{1}[engagement(i)] \quad .$$
\end{enumerate}

\subsection*{Learning the Ranking Function}
Given the user-item-tweet features $\phi(u,i,d)$ extracted based on user-item-tweet interactions and the corresponding metadata of the tweets our goal is to use the observed engagement for each triple to optimize the parameters of the ranking function for the target information retrieval metric, in our case, nDCG@10. 
In this work we choose to use an ensemble of MART and LambdaMART learning-to-rank methods~\cite{export:132652}. In particular, LambdaMART was chosen due to its the ability to directly optimize the Information Retrieval (IR) metric nDCG@10 and the performance exhibited in recent ranking competitions~\cite{DBLP:journals/jmlr/BurgesSBPW11,DBLP:journals/jmlr/ChapelleC11}. 

LambdaMART uses as base learners Multiple Additive Regression Trees or MART, also known as Gradient Boosted Regression Trees (GBRT)~\cite{friedman2001, Friedman:2002:SGB:635939.635941}. LambdaMART is a pairwise learning-to-rank method capable to optimize arbitrary IR metrics by guiding the learning process using so-called $\lambda$-gradients, which reflect small changes in the IR metric while iterating over the training set. We omit a detailed description of the model, since it is out of the scope of this report, and refer the reader to~\cite{export:132652} and~\cite{DBLP:journals/jmlr/BurgesSBPW11} for a comprehensive analysis of LambdaMART.
\section{Experiments}
\label{sec:experiments}
In this section, we first present some key statistics of the dataset used in the experiments, followed by the experimental evaluation of our approach.
\subsection*{Dataset}
The dataset used in our experiments corresponds to the \emph{MovieTweetings Extended}, released as part of the RecSys Challenge~\cite{Dooms13crowdrec,2014_recsys_challenge_1,2014_recsys_challenge_2}, 2014. For this challenge, the dataset is chronologically split in three subsets: training, test, and evaluation set, whose statistics are summarized in Table~\ref{tab:dataset_stats}. Note that the number of \emph{unique} users and items (considering the ones present in training and test) are |U| = 23,555  and |I| = 14,316, respectively. 

\begin{figure}[!tb]
\centering
\includegraphics[width=\columnwidth,clip]{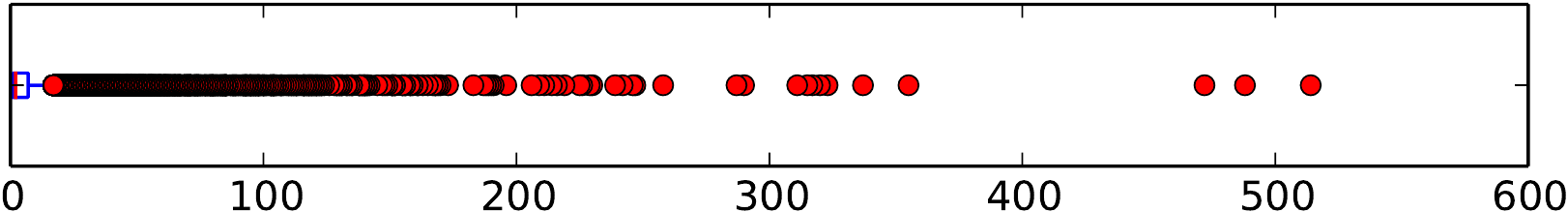}
{number of interactions per user}
\caption{Boxplot of the distribution of the number of interactions (i.e., user-item-tweet triples) per user in the training set. There is a minimum number of 1 and a maximum of 514 interactions per user, with mean value of 7.71 and a median of 2.}
\label{fig:boxplot_training}
\end{figure}

Figure~\ref{fig:boxplot_training} shows a boxplot of the distribution of the number of interactions (i.e., user-item-tweet triples) per user in the training set. Note that the number of interactions is very low for most users, leading to a median of 2. We can also observe many outliers, e.g., users with over 300 interactions, far from them mean of 7.71.

\begin{table}[!tb]
\centering
\begin{tabular}{ l  r  r  r  p{2.5cm} }
\toprule
\textbf{Split} & \textbf{Users} & \textbf{Items} & \textbf{Tweets} & \textbf{Tweets creation} \\
\midrule
Training & 22,079 & 13,618  & 170,285 & 2013-02-28 14:43 - 2014-01-08 22:06 \\
Test     & 5,717  & 4,226   & 21,285  & 2014-01-08 22:06 - 2014-02-11 15:49 \\
Evaluation & 5,514  & 4,559 & 21,287  & 2014-02-11 15:49 - 2014-03-24 9:57 \\
\midrule
Total & 24,924 & 15,142 & 212,857 & 2013-02-28 14:43 - 2014-03-24 9:57 \\
\bottomrule
\end{tabular}
\caption{MovieTweetings Extended dataset statistics.}
\label{tab:dataset_stats}
\end{table}

The training set consists of the first 80\% tweets and is used as input for our collaborative ranking approach. The test and evaluation sets both contain 10\% of the remaining tweets. The test split is used to measure the predictive performance of our method, and the final evaluation split is reserved by the challenge organizers for the final assessment.

\subsection*{Evaluation Methodology}
\subsubsection*{Metric}
The evaluation for the task of predicting which user-item-tweet triples generate the highest user engagement is measure by the IR metric Normalized discounted cumulative gain with a cut level equal to 10 (nDCG@10), nDCG  measures the performance of a recommendation system based on the graded relevance of the ranked tweets based on their predicted engagement. It varies from 0.0 to 1.0, with 1.0 representing the ideal ranking of the tweets. This metric is commonly used in information retrieval and to evaluate the performance of web search engines~\cite{2011_ricBaeza}. 

Formally, the discounted cumulative gain (DCG) is a measure of ranking quality defined as:
\begin{equation}
DCG@K = \sum_{k=1}^K \frac{2^{rel_k} - 1}{\log_2 (k + 1)}
\label{eq:dcg_at_k}
\end{equation}

if we define $IDCG@K$ as the maximum possible (ideal) DCG for a given set of user-item-twitter triples and engagement, we can normalize the DCG@K to be between 0.0 to 1.0 and obtain the nDCG@K :
\begin{equation}
nDCG@K = \frac{DCG@K}{IDCG@K}
\label{eq:ndcg_at_k}
\end{equation}
where $K$ is the maximum number of tweets that are considered in the ranked list for the evaluation.

For each user in the test set we compute the corresponding nDCG@10 and then average over all users. For the evaluation, we use the tool provided by the organizers, which is based on RiVal, a toolkit for recommender systems evaluation.\footnote{\textbf{RiVal}: \url{https://github.com/recommenders/rival}} All results reported in this section corresponds to nDCG@10 obtain in the test split only.

\subsubsection*{Baselines}
We compare the performance of our collaborative ranking approach for user engagement prediction, or \emph{CRUE}, to the baselines recommender systems described next.

\begin{itemize} 
\item \textbf{FM} a Factorization Machine~\cite{rendle:tist2012}. We use the FM reference implementation provided by libFM\footnote{\textbf{libFM}: \url{www.libfm.org}} and train the model for a ranking task using Markov Chain Monte Carlo (MCMC), and found that a standard deviation of 0.1, 16 factors, and 1000 iterations, lead to good results. We include the indicator variables corresponding to users, items, and ranting, and the engagement as label, which is equivalent to a pairwise tensor factorization for ranking~\cite{rendle:tist2012}.

\item \textbf{recRating} predicts the rating associated to the triple user-item-tweet as the user engagement value. 
\item \textbf{recHEI} predicts as future engagement the average value of the user engagement that a given item received in the past.
\item \textbf{recRandom} predicts a random value for the user engagement.
\end{itemize}
Observe that recRating and recHEI correspond to features F1 and F8, respectively, as defined in Section~\ref{sec:methods}.

\subsubsection*{CRUE Setup}
As discussed in Section~\ref{sec:methods}, CRUE uses a linear ensemble of MART and LambdaMART to learn the ranking function for user engagement prediction. LambdaMART uses as base learners Multiple Additive Regression Trees or MART, also known as Gradient Boosted Regression Trees (GBRT)~\cite{friedman2001, Friedman:2002:SGB:635939.635941}, whose main parameters to adjust is the number of estimators, the number of trees in the forest, and the size of the random subsets of features to consider when splitting a node.

In order to reduce noise in the training dataset (cf., Figure~\ref{fig:boxplot_training}), we drop the user interactions of users with less than 4 and over 200 interactions, and use the renaming users for training. We observe more stable results after this pruning.

We set the number of trees using an additional 80\%-20\% partition of the training set, keeping the 20\% as a validation split to determine the optimal number of trees. Based on the same validation procedure, we found that a number of leaves for each tree equal to 10 and a learning rate (or shrinkage) equal to 0.1 gave us good results.

We perform early stopping in the training when we observed that no additional improvement, in terms of nDCG@10, was achieved in the validation split after 50 consecutive rounds. We set the maximum number of features for splitting equal to the number of features in the training set.

To implement CRUE we use a set of Python scripts that rely upon NumPy, SciPy\footnote{\textbf{NumPy and SciPy}:  \url{http://www.scipy.org/}} and scikit-learn~\cite{scikit-learn}. We use RankLib's implementation of LambdaMART to learn the ranking function and directly optimOur goal is to reduce noise and wize for nDCG@10~\cite{2014_lemur_ranklib}.

\subsection*{Results}
\begin{figure}[!tb]
\centering
\includegraphics[width=\columnwidth,clip]{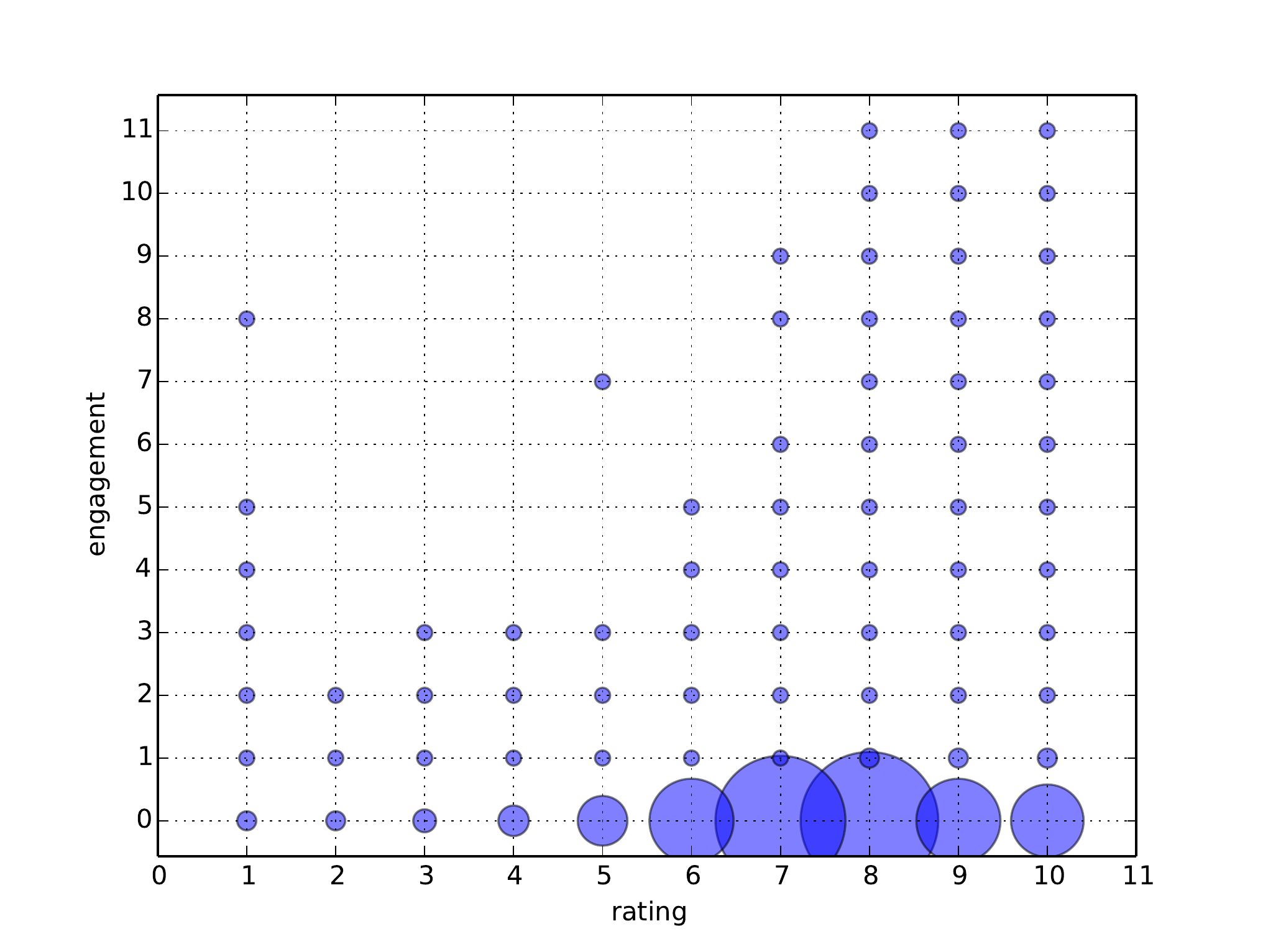}
\caption{Scatter plot in (rating, engagement) coordinates, where the area of the circle depicts the frequency observed for the corresponding pair, i.e., the larger the radius, the higher the frequency.}
\label{fig:rating_engagement}
\end{figure}

Table~\ref{tab:results} summarizes the best results obtained by each method in our experiments. We found that the explicit feedback expressed as rating is quite powerful predictor for user engagement, this is the only source of information for the simple recommender recRating that achieves a nDCG@10 of 0.8182, outperforming the factorization model FM. Figure~\ref{fig:rating_engagement} shows a scatter plot depicting the global rating-engagement relationship in the training set, we can observe that higher ratings tend to receive higher engagement.

The historical user engagement that an item received in the past is also a very good predictor for the task at hand, recHEI also outperforms the FM, but is not as good as recRating.

We also experiment training a FM with additional contextual features, e.g., the ones extracted from tweet metadata, as we did for CRUE, but the performance obtained in our experiments was below to the one of using the user-item-rating triples alone. We are still engineering additional features to train a new FM in order to include the model in a potential ensemble together with CRUE.

CRUE outperforms the baselines considered here and, at the time of writing, our approach places our team (SUD) within the top-5 in the RecSys Challenge leaderboard~\cite{2014_recsys_challenge}. 

\begin{table}[!tb]
\centering
\begin{tabular}{ l  c}
\toprule
\textbf{Recommender} & \textbf{nDCG@10} \\
\midrule
CRUE (our approach) & \textbf{0.8701}\\
FM & 0.8023 \\
recRating & 0.8182\\
recHEI & 0.8031\\
recRandom &  0.7532\\
\bottomrule
\end{tabular}
\caption{Results.}
\label{tab:results}
\end{table}

\paragraph{Some general thoughts}
\begin{itemize}
\item During the competition we concentrated first in directly optimizing the IR metric used in the challenge, namely nDCG@10. The results of our collaborative ranking approach, CRUE, are the ones reported in this paper. Our goal towards the end of the challenge is to continue trying different models, which can explain the variability of the data.

\item We found, in initial experiments, that factorization models that use the engagement as label, either for regression or classification, do not perform as good as the rating (recRating) or the item average engagement (recHEI).

\item We observe that smoothing of the features, given the long tail distribution of most of them, works well. We smooth them using square root of the values, as shown in Section~\ref{sec:methods}.

\item Normalization of the features using their z-scores also helps us to significantly improve ranking performance.

\item Removing user outliers also helps.

\item An initial exploration on the individual nDCG@10 performance per user indicates that for some users are rather \emph{easy} to predict the engagement that their tweets will receive. Training models on \emph{hard} users only, e.g., those whose nDCG@10 is less than a threshold, is equally effective than training using all of the user available in the training set. This idea goes towards selective sampling and active learning, which have been successfully applied in the context of CF~(e.g., \cite{Diaz-Aviles:2012:RTR:2365952.2365968}).

\item In our experiments, we found that traditional kNN and Matrix factorization models for CF did not achieve a performance better than using the rating as predictor for user engagement. We are still trying to improve the performance of these approaches to include them in a ensemble together with CRUE.

\item We also collected complementary information from IMDb in order to enrich the movie profiles, e.g., we collected the year of release of the movie, the description, tags, as well as the director and actors. We found that in some cases the year of release improve marginally our predictions. Note that the results reported in this paper are based only on the dataset provided.

\item It would be very interesting to consider a different evaluation protocol that better reflects the user impact in the engagement that her tweets will receive when compared to other users as well. In other words, assess the user engagement that triples user-item-tweet from user $u$ will receive over other user $v$'s triples.

\item In independent experiments, we have identified that the time of the retweets can be very useful to predict engagement at different levels, e.g., userwise or topicwise. Having access to this additional feature could also help us boost the results.

\item If an application requires the user engagement to be predicted on-the-fly, it can leverage the explicit feedback (rating) and the historic user engagement per item for an effective and fast prediction.
\end{itemize}
\section{Related Work}
\label{sec:related_work}
The popularity of Twitter, easy access to data and the unique characteristic of velocity and real-time information propagation, have made it an attractive domain recently. For instance, in \cite{ICWSM112790} the authors make an effort on characterizing retweeters and their retweeting behavior, which is related to our goal of predicting user engagement. However, existing work does not consider a collaborative ranking approach to this task, as the one introduced in this paper. 

Our approach is also related to context-aware recommender systems~\cite{2011_recsys_handbook}, which use additional contextual information such as time, location, or the company of other people to improve recommendation performance. Our approach leverages features extracted from the rich metadata available for the tweets and follow a collaborative ranking approach, which is different to the state-of-the-art approaches for context aware recommender systems that primarily rely on factorization methods~\cite{rendle:tist2012,Karatzoglou:2010:MRN:1864708.1864727}. 

Existing collaborative ranking approaches include the ones proposed in~\cite{Diaz-Aviles:2012:SRR:2365952.2366001} and \cite{Balakrishnan:2012:CR:2124295.2124314}, these methods also cast the recommendation problem as a learning-to-rank task, but in the absence of rich features, they use factors output by a matrix factorization algorithm for CF as feature vectors, and then learn a ranking function. We instead use the rich metadata from tweets to build feature vectors that characterize the user-item-tweet interactions. As future work, we plan to also include the latent factors as part of our existing feature vectors to asses to what extent we can improve the personalized ranking. Another interesting collaborative filtering approach is the one introduced in ~\cite{NIPS2012_4829}, which constructs the feature vectors based on metrics derived from the user neighborhoods.

None of the existing approaches have explored the collaborative ranking setting for user engagement prediction as we do in this work.
\section{Conclusion and Future Work}
\label{sec:conclusion}
In this work we showed how our collaborative ranking approach is able to predict user engagement in Twitter. The rich metadata available from Twitter API help us to build feature vectors that characterize the user-item-tweet interactions, and complements the information captured by the explicit feedback (i.e., rating) expressed in the tweet. 

Experimental results show that by using our collaborative ranking approach is possible to predict attractive tweets that will engage more people. Such insights can be then leveraged, for example, to make interesting predictions for business success~\cite{10.1109/ASONAM.2010.50} or public safety~\cite{ICWSM124594}. 

We are currently investigating additional features such as characteristics of the user and item social neighborhoods to improve ranking performance. An additional direction for future work is the extension of our model for streaming data scenarios, e.g.,~\cite{Diaz-Aviles:2012:RTR:2365952.2365968}, where the model parameters are learned online without compromising ranking performance.

\bibliographystyle{abbrv} 
\bibliography{2014_recsys_challenge_6_sud_diaz-aviles_biblio}
\end{document}